\documentclass[epj,nopacs]{svjour}
\usepackage{graphics}
\begin{document}
\title{Hadron spectra, flow and correlations in PbPb collisions at the
LHC: interplay between soft and hard physics}
%\subtitle{Do you have a subtitle?\\ If so, write it here}
\author{I.P. Lokhtin\inst{1}, A.V. Belyaev\inst{1}, L.V. Malinina\inst{1,2}, S.V. Petrushanko\inst{1}, E.P. Rogochaya\inst{2} \and A.M. Snigirev\inst{1} % etc
% \thanks is optional - remove next line if not needed
%\thanks{\emph{Present address:} Insert the address here if needed}%
}                     % Do not remove
%
%\offprints{}          % Insert a name or remove this line
%
\institute{Skobeltsyn Institute of Nuclear Physics, Lomonosov Moscow State University, Moscow, Russia \and Joint Institute for Nuclear Research, Dubna, Russia}

%
%\date{Received: date / Revised version: date}
% The correct dates will be entered by Springer
%
\abstract{The started LHC heavy ion program makes it possible to
probe new frontiers of the high temperature Quantum
Chromodynamics. It is expected that the role of hard and semi-hard
particle production processes may be significant at ultra-high
energies even for bulk properties of the created matter. In this
paper, the LHC data on multiplicity, hadron spectra,
elliptic flow and femtoscopic correlations from PbPb collisions at
center-of-mass energy 2.76 TeV per nucleon pair are analyzed in
the framework of the HYDJET++ model. The influence of the jet
production mechanism on these observables is discussed.
%
%\PACS{
%      {12.38.Mh}{Quark-gluon plasma} \and
%      {25.75.-q}{Relativistic heavy-ion collisions} \and
%      {25.75.Bh}{Hard scattering in relativistic heavy ion collisions}
%     } % end of PACS codes
}
%end of abstract
%
\titlerunning{Hadron spectra, flow and correlations in PbPb collisions at the
LHC...}
\authorrunning{I.P. Lokhtin et al.}
\maketitle
\section{Introduction}
One of the main aims of modern high energy physics is to study the
strong interaction in extreme regimes at
super-high densities and temperatures. Hot and dense multi-parton
and multi-hadron systems are produced in high-energy nuclear
collisions and can be investigated within the Quantum
Chromodynamics (QCD) (see, e.g., recent
reviews~\cite{d'Enterria:2006su,BraunMunzinger:2007zz,Salgado:2009jp,Dremin:2010jx}).
The experimental and phenomenological study of multi-particle
production in ultrarelativistic heavy ion collisions is expected
to provide valuable information on the dynamical behavior of QCD
matter in the form of the quark-gluon matter (QGM), as predicted by
lattice calculations. The current lead-lead (PbPb) collision 
center-of-mass energy per nucleon pair at LHC, $\sqrt s_{\rm NN}=2.76$ TeV, 
is a factor of $\sim 14$ larger than that at RHIC and makes it possible to 
probe new frontiers of the super-high temperature and (almost) net-baryon
free QCD. It is expected that a role of hard and semi-hard 
particle production may be significant at ultra-high energies 
even for bulk properties of the created matter.

A number of striking LHC results from PbPb runs have been 
published by
ALICE~\cite{Aamodt:2010pb,Aamodt:2010pa,Aamodt:2010jd,Aamodt:2010cz,Aamodt:2011mr,ALICE:2011ab,Aamodt:2011by,Aamodt:2011vg,Abelev:2012ej,Abelev:2012rv,Abelev:2012nj,ALICE:2012aa,ALICE:2012di},
ATLAS~\cite{Aad:2010bu,Aad:2010px,ATLAS:2011yk,ATLAS:2011ag,Aad:2012bu}
and CMS
~\cite{Chatrchyan:2011sx,Chatrchyan:2011ua,Chatrchyan:2011ek,Chatrchyan:2011pe,Chatrchyan:2011pb,Chatrchyan:2012vq,CMS:2012aa,Chatrchyan:2012np,Chatrchyan:2012ni,Chatrchyan:2012wg,Chatrchyan:2012ta,Chatrchyan:2012xq,Chatrchyan:2012gt,Chatrchyan:2012mb}
collaborations (see~\cite{Muller:2012zq} for the overview of the results from the first year of heavy 
ion physics at LHC). In the paper, we analyze the LHC data on
multiplicity, hadron spectra, elliptic flow and femtoscopic
momentum correlations from PbPb collisions at $\sqrt s_{\rm NN}=2.76$ TeV in
the framework of the HYDJET++ model~\cite{Lokhtin:2008xi}. The
main goal of this study is to show a significant influence of the
(in-medium) jet production mechanism on the global event pattern
at LHC. Note that the influence of the interactions between jets and 
soft hadrons on a number of physical observables at the LHC was studied  
recently in the framework of the EPOS model~\cite{Werner:2012xh}. 

\section{HYDJET++ model}

HYDJET++ (the successor of HYDJET~\cite{Lokhtin:2005px}) is the
model of relativistic heavy ion collisions, which includes two
independent components: the soft state (hydro-type) and the hard
state resulting from the in-medium multi-parton fragmentation. The
details of the used physics model and simulation procedure can be
found in the HYDJET++ manual~\cite{Lokhtin:2008xi}. Below main
features of the model are described only very briefly.

\subsection{Soft component}

The soft component of an event in HYDJET++ is the ``thermal''
hadronic state generated on the chemical and thermal freeze-out
hypersurfaces obtained from the pa\-ra\-met\-ri\-za\-ti\-on of
relativistic hydrodynamics with preset freeze-out conditions (the
adapted event generator FAST
MC~\cite{Amelin:2006qe,Amelin:2007ic}). Hadron multiplicities are
calculated using the effective thermal volume approximation and
Poisson multiplicity distribution around its mean value, which is
supposed to be proportional to a number of participating nucleons
for a given impact parameter of a AA collision. To simulate the
elliptic flow effect, the hydro-inspired parametrization is
implemented for the momentum and spatial anisotropy of a soft
hadron emission source~\cite{Lokhtin:2008xi,Wiedemann:1997cr}.

The soft hadron simulation procedure includes the following steps.
\begin{itemize}
\item Generation of a hadron 4-momentum in the liquid element
rest frame in accordance with the equilibrium distribution
function.
\item Generation of a spatial position and local 4-velocity of the
liquid element in accordance with the phase space and the
character of fluid motion.
\item Standard von Neumann rejection/acceptance procedure to
account for a difference between the true and generated
probabilities.
\item Boost of the hadron 4-momentum in the event center-of-mass
frame.
\item Taking into account two- and three-body decays of resonances
with branching ratios taken from the SHARE particle decay
table~\cite{Torrieri:2004zz}.
\end{itemize}
The model has a number of input parameters for the soft component.
They are tuned from fitting to experimental data values for various
physical observables (see the next section). Parameters which
determine the hadron ratios are the chemical freeze-out
temperature $T_{\rm ch}$ and thermodynamical potentials (baryon
potential $\mu_{\rm B}$, strangeness potential $\mu_{\rm S}$,
isospin potential $\mu_{\rm Q}$, and potential of positively
charged pions $\mu_{\pi}^{\rm th}$ at thermal freeze-out). We have
used $T_{\rm ch}=165$ MeV and, for simplicity, zero
thermodynamical potentials. The collective flow parameters are the
maximal longitudinal and transverse flow rapidities, $Y_{\rm
L}^{\rm max}=4.5$ and $Y_{\rm T}^{\rm max}=1.265$, respectively.
The latter value has been fixed from the data on hadron
$p_{\rm T}$-spectra together with the thermal freeze-out temperature
$T_{\rm th}=105$ MeV. The total multiplicity of the soft component
is determined by space-time parameters on the thermal freeze-out
for central collisions, maximal transverse radius $R_{\rm f}(b=0)$,
proper time $\tau_{\rm f}(b=0)$ and the duration of hadron emission
$\Delta \tau_{\rm f}(b=0)$. These parameters have been fixed from
fitting of three-dimensional correlation functions measured for
$\pi^+\pi^+$ pairs and extracting the correlation radii:
$R_{\rm f}(b=0)=13.45$ fm, $\tau_{\rm f}(b=0)=12.2$ fm/$c$, $\Delta
\tau_{\rm f}(b=0)=2$ fm/$c$.

\subsection{Hard component}

The model used for the hard component in HYDJET++ is the same as
in the HYDJET event generator. It is based on the PYQUEN partonic
energy loss model~\cite{Lokhtin:2005px}. The approach describing
the multiple scattering of hard partons is based on accumulated
energy loss via gluon radiation which is associated with each
parton scattering in expanding quark-gluon fluid. It also includes
the interference effect in gluon emission with a finite formation
time using the modified radiation spectrum $dE/dx$ as a function
of the decreasing temperature $T$. The model takes into account
radiative and collisional energy loss of hard partons in
longitudinally expanding quark-gluon fluid, as well as the
realistic nuclear geometry. Radiative energy loss is treated in
the framework of the BDMS model~\cite{Baier:1999ds,Baier:2001qw}
with a simple generalization to a massive quark case using the
``dead-cone'' approximation~\cite{Dokshitzer:2001zm}. The
collisional energy loss due to elastic scatterings is calculated
in the high-momentum transfer
limit~\cite{Bjorken:1982tu,Braaten:1991jj,Lokhtin:2000wm}. The
strength of the energy loss in PYQUEN is determined mainly by the
initial maximal temperature $T_0^{\rm max}$ of hot matter in
central PbPb collisions (i.e. an initial temperature in the center
of the nuclear overlap at mid-rapidity). The energy loss depends
also on the proper time $\tau_0$ of QGM formation and the number
$N_{\rm f}$ of active flavors in the medium. The parameter values
$T_0^{\rm max}=1$ GeV, $\tau_0=0.1$ fm$/c$ and $N_{\rm f}=0$
(gluon-dominated plasma) are used at $\sqrt s_{\rm NN}=2.76$ TeV.
Another important ingredient of PYQUEN is the angular spectrum of
in-medium radiation. Since the ATLAS~\cite{Aad:2010bu} and
CMS~\cite{Chatrchyan:2011sx} data on the dijet asymmetry support
the supposition of the ``out-of-cone'' medium-induced partonic
energy loss, the ``wide-angular'' pa\-ra\-met\-ri\-za\-ti\-on of
medium-induced gluon radiation~\cite{Lokhtin:2011qq} is used in
the presented investigation.

The simulation of single hard nucleon-nucleon sub-collisions by
PYQUEN is constructed as a modification of the jet event obtained
with the generator of hadron-hadron interactions
PYTHIA$\_$6.4~\cite{Sjostrand:2006za}. We used PYTHIA tune
Pro-Q20 for the present simulation. It reproduces the CMS data on
hadron momentum spectra in pp collisions at LHC energies with the
10--15\% accuracy in the full measured
$p_{\rm T}$-range~\cite{Chatrchyan:2011av}.

The event-by-event simulation procedure in PYQUEN includes the following steps.
\begin{itemize}
\item Generation of initial parton spectra with PYTHIA and
production vertices at a given impact parameter.
\item Rescattering-by-rescattering simulation of the parton path
in a dense zone and its radiative and collisional energy loss.
\item Final hadronization according to Lund string model for
hard partons and in-medium emitted gluons.
\end{itemize}
Then the number of PYQUEN jets is generated according to the
binomial distribution. The mean number of jets produced in an AA
event is calculated as a product of the number of binary NN
sub-collisions at a given impact parameter per the integral cross
section of the hard process in NN collisions with the minimum
transverse momentum transfer $p_{\rm T}^{\rm min}$. The latter is an
input parameter of the model. In the framework of HYDJET, partons
produced in (semi)hard processes with the momentum transfer lower
than $p_{\rm T}^{\rm min}$, are considered as being ``thermalized''. So,
their hadronization products are included ``automatically'' in the
soft component of the event. In order to take into account the
effect of nuclear shadowing on parton distribution functions, we
use the impact parameter dependent pa\-ra\-met\-ri\-za\-ti\-on
obtained in the framework of Glauber-Gribov
theory~\cite{Tywoniuk:2007xy}. The nuclear shadowing correction
factor $S(r_1, r_2, x_1, x_2, Q^2)$ depends on kinematic variables
of incoming hard partons (the momentum fractions $x_{1,2}$ of
initial partons from incoming nuclei and the momentum scale $Q^2$
of a hard scattering), and on the transverse coordinates $r_{1,2}$
of partons in their parent nuclei. This initial state effect
reduces the number of partons in the incoming hadronic
wave-functions of both nuclei and, thus, reduces the yield of
high-$p_{\rm T}$ hadrons coming from jets (in addition to the
medium-induced partonic energy loss).

To study the femtoscopic (HBT) momentum correlations, it is
necessary to specify a space-time structure of a hadron emission
source. The information about final particle 4-momenta and
4-coordinates of emission points is automatically recorded using
the soft hadron simulation procedure (see previous subsection). 
For the hard component, the existence of a ``source'' with a given size is
not obvious and is a matter of discussion (see, for instance,
\cite{Paic:2005cx}). The point of jet hadronization
and the point of initial parton-parton hard scattering do not
coincide. In accordance with~\cite{Paic:2005cx}, we assume
that hadronization occurs at different distances from an initial 
hard scattering, depending on an energy of a jet.

Our treatment of the coordinate information for low momentum jet particles 
with $p_{\rm T}<1$ GeV/$c$ is similar to one for directly produced 
hadrons from soft component. Such particles are emitted from the 
fireball of radius $R_{\rm f}$ at mean proper time $\tau_{\rm f}$ 
with the emission duration $\Delta \tau_{\rm f}$.

Four-coordinates of high momentum jet particles with $p_{\rm T}>1$ GeV/$c$ are shifted from those of a jet vertex and coded in the following 
way~\cite{Paic:2005cx}:
\begin{eqnarray}
& & x_i=x_0+L_{li}\frac{\cos \phi_j}{\cosh y_j}+L_{t1i}\frac{\cos \phi_j \sinh y_j}{\cosh y_j}-L_{t2i}\sin \phi_j\nonumber\\
& & y_i=y_0+L_{li}\frac{\sin \phi_j}{\cosh y_j}+L_{t1i}\frac{\sin \phi_j\sinh y_j}{\cosh y_j}+L_{t2i}\cos \phi_j\nonumber\\
& & z_i=z_0+L_{li}\frac{\sinh y_j}{\cosh y_j}-L_{t1i}\frac{1}{\cosh y_j} \nonumber\\
& & t_i=t_0+L_{li}~,
\end{eqnarray}
where the subscripts $i$ and $j$ refer to particles and jets respectively.
Here $x_0=r\cos\psi$ and $y_0=r\sin\psi$ are transverse
coordinates of a jet production vertex, $r$ is a distance from the
nuclear collision axis to the vertex with the azimuthal angle
$\psi$. The longitudinal position and the initial time of a hard
process are estimated as $z_0\simeq \sinh y_j/p_{{\rm T}j}$ and 
$t_0\simeq \cosh y_j/p_{{\rm T}j}$,
respectively ($\phi_j$ and $y_j$ are the azimuthal angle and the
pseudorapidity of a jet, respectively). For each hadron
$i$, the localization $L_{li}$ of hadronization along the
direction of a jet is randomized from Gaussian
distributions with mean $l_i$ and variance
$\sigma=l_i/3$~\cite{Paic:2005cx}. The hadronization points
$L_{t1i}$ and $L_{t2i}$ (with respect to the jet direction) are
randomized in the transverse direction from a Gaussian distribution with
variance $\sigma_t=0.5$~fm and zero mean value; $l_i$ is assumed
to be proportional to jet momentum $|{\bf p}_j|$. Thus, $l_i
=f|{\bf p}_j|$ where $f$ is a multiplicative factor that
represents our lack of theoretical insight into the process of
hadronization (the value $f=5$ being used in our simulations).
For simplicity, the jet axis is determined by the momentum 
direction of a leading particle in the event. Then four-coordinates of jet particles with $p_{\rm T}>1$ GeV/$c$ are shifted from the jet vertex position along or opposite this direction depending on a sign of the scalar production $(\bf {p}_i \bf{p}_j)$. 

\section{Numerical results and discussion}

It was demonstrated in~\cite{Lokhtin:2008xi} that the HYDJET++
model can describe bulk properties of the hadronic state created
in AuAu collisions at RHIC at $\sqrt s_{\rm NN}=200$ GeV (hadron
spectra and ratios, radial and elliptic flow, femtoscopic momentum
correlations), as well as the high-$p_{\rm T}$ hadron spectra. At
present we apply HYDJET++ with tuned input parameters to reproduce
the LHC data from PbPb collisions, and to estimate an
influence of the hard production mechanism on physics observables.

\subsection{Centrality and pseudorapidity dependence of multiplicity}

\begin{figure*}
%\begin{center}
\resizebox{0.5\textwidth}{!}{%
\includegraphics{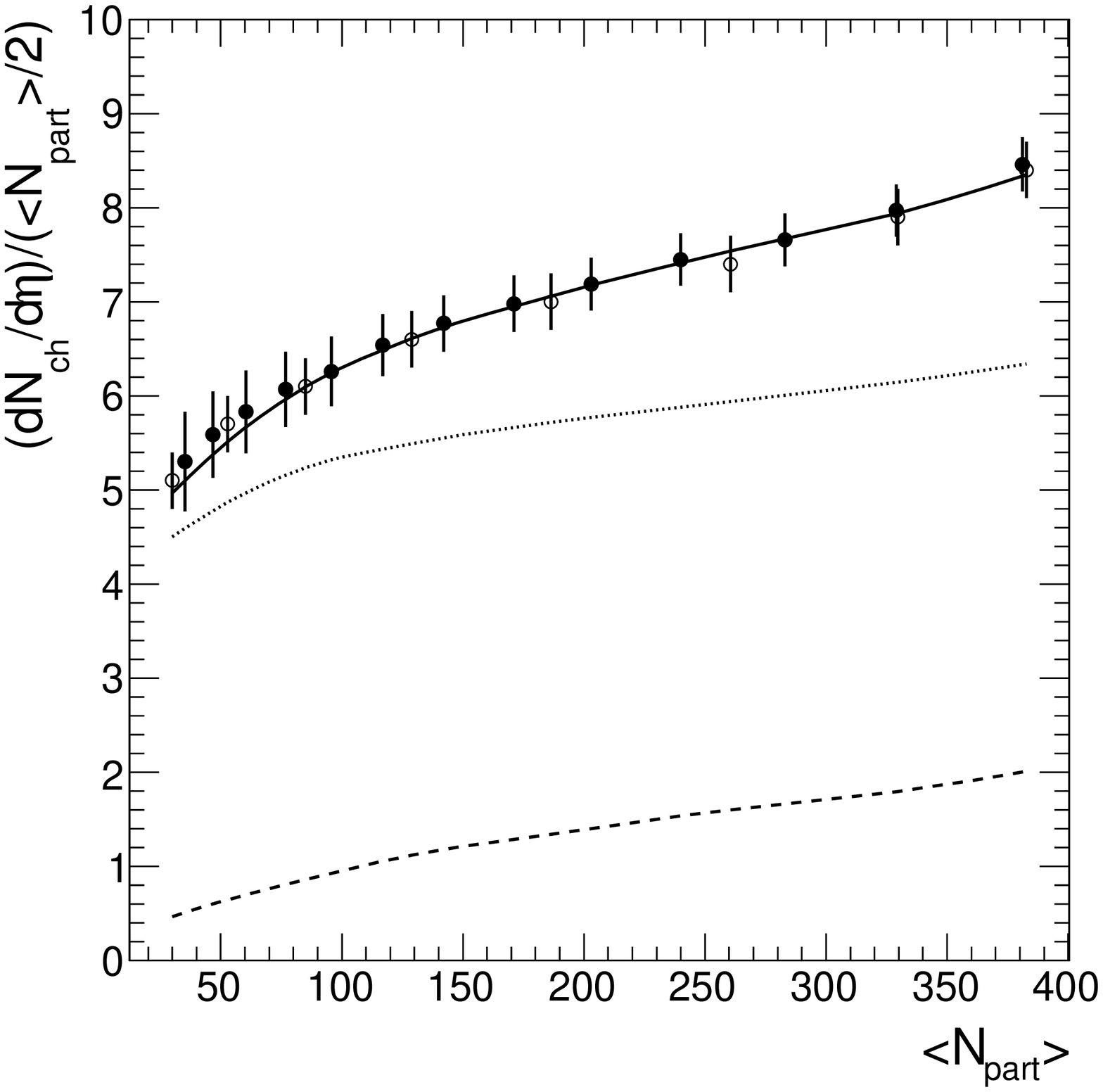}}
\resizebox{0.5\textwidth}{!}{%
\includegraphics{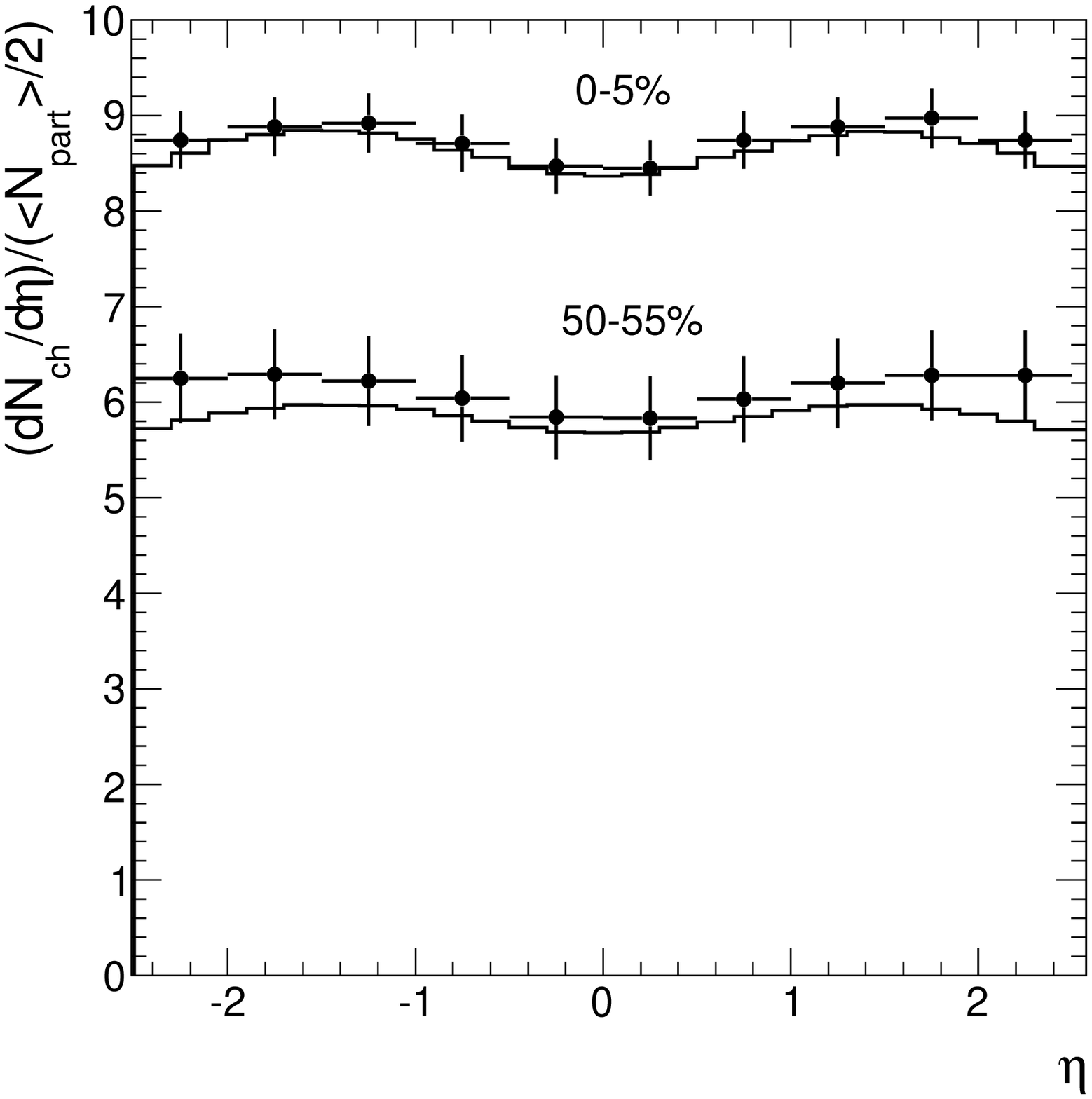}}
%\end{center}
 \caption{Charged multiplicity density at mid-rapidity (normalized
on the mean number of participating nucleons $\left<N_{\rm
part}\right>$) as a function of PbPb event centrality (left) and
pseudorapidity distribution in two centrality bins (right) 
at $\sqrt s_{\rm NN}=2.76$ TeV. The open and closed points are 
ALICE~\cite{Aamodt:2010jd} and CMS~\cite{Chatrchyan:2011pb} data respectively, 
curves (left) and histograms (right) are the simulated HYDJET++ events (solid --
total result, dashed -- hard component, dotted -- soft component).} 
\label{fig1}
\end{figure*}

Figure~\ref{fig1} shows the charged multiplicity density at
mid-rapidity (normalized on the mean number of participating
nucleons $\left<N_{\rm part}\right>$) as a function of event 
centrality (left) and pseudorapidity distribution in two  
centrality bins (right) in PbPb collisions at $\sqrt s_{\rm NN}=2.76$ TeV. The
results of HYDJET++ simulation are compared with the ALICE~\cite{Aamodt:2010jd} 
and CMS~\cite{Chatrchyan:2011pb} data. Since multiplicities from soft and hard 
components depend on the centrality in a different way, the relative contribution 
of soft and hard components to the total event multiplicity can be fixed 
through the centrality dependence of $dN/d\eta$. The latter can be reproduced at $\sim
25$\% contribution of the hard component (versus $\sim 15$\% at
RHIC~\cite{Lokhtin:2008xi}) in most central collisions, and
it decreases for more peripheral events. This contribution of the
hard component corresponds to the minimal transverse momentum
transfer of parton-parton scatterings $p_{\rm T}^{\rm min}=8.2$ GeV/$c$.
The approximately flat $\eta$ dependence of the multiplicity observed by 
CMS (the variation being less than 10\% in the 
pseudorapidity range $|\eta|<2.5$) is also well reproduced by HYDJET++ 
simulation for different event centralities at reasonable values of maximal 
longitudinal flow rapidity, $Y_{\rm L}^{\rm max}=3.5\div 4.5$. 

The ratios of mid-rapidity
yields of final charged had\-rons in the model (excluding weak
decays of strange had\-rons) are $K^{\pm}/\pi^{\pm}=0.153$ and
$p^{\pm}/\pi^{\pm}=0.065$ at 0--5\% of PbPb centrality. The
calculated kaon-to-pion ratio reproduces the experimental number
reported by ALICE~\cite{Floris:2011ru} ($0.155 \pm 0.012$), while
the proton-to-pion ratio is overestimated in the model by $\sim
40$\% as compared with the ALICE data ($0.046 \pm 0.004$). Since
pure thermal models meet with difficulties to reproduce these hadron
ratios, in our opinion, it may be interpreted as a significant
influence of (mini-)jet production on a particle composition in
heavy ion collisions (hadron ratios for soft and hard components
are quite different).

\subsection{Transverse momentum spectrum and nuclear modification factor}

\begin{figure*}
%\begin{center}
\resizebox{0.5\textwidth}{!}{%
\includegraphics{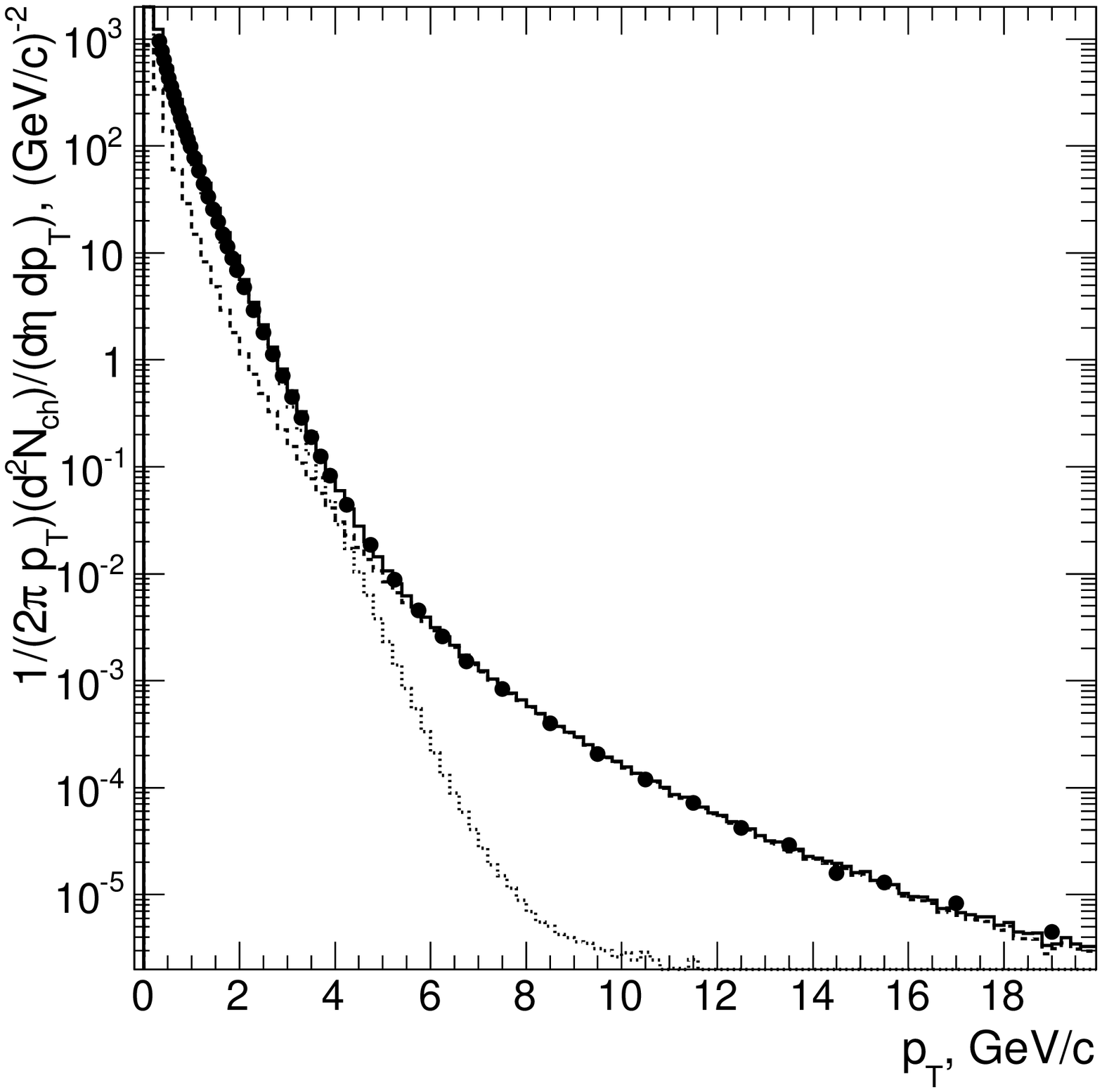}}
\resizebox{0.5\textwidth}{!}{%
\includegraphics{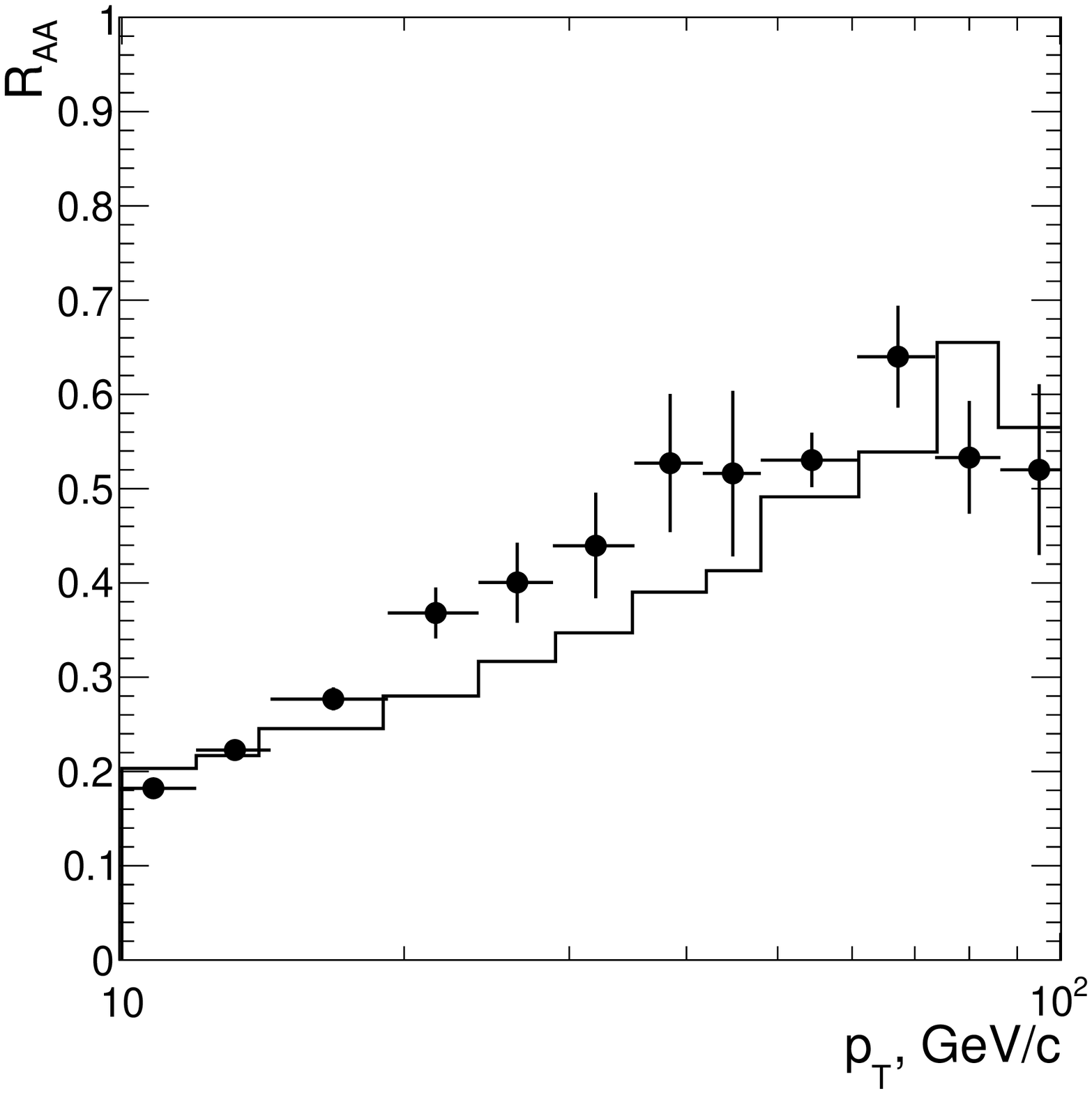}}
%\end{center}
 \caption{Charge particle transverse momentum spectrum (left) and 
nuclear modification factor $R_{\rm AA}(p_{\rm T})$ (right) in 5\% of most 
central PbPb collisions at $\sqrt s_{\rm NN}=2.76$ TeV. The points 
are ALICE~\cite{Aamodt:2010cz} (left) and CMS~\cite{CMS:2012aa} 
(right) data, histograms are the simulated HYDJET++ events (solid --
total result, dashed -- hard component, dotted -- soft component).} 
\label{fig2}
\end{figure*}

Figure~\ref{fig2} (left) shows the transverse momentum spectrum of charged 
particles ($|\eta|<0.8$) in 5\% of most central PbPb collisions at 
$\sqrt s_{\rm NN}=2.76$ TeV. The results of HYDJET++ simulation are compared 
with the ALICE data~\cite{Aamodt:2010cz}. The slope of $p_{\rm T}$-spectra allows one 
to fix the thermal freeze-out temperature $T^{\rm th}=105$ MeV and
the maximal transverse flow rapidity $Y_{\rm T}^{\rm max}=1.265$.
One can see that HYDJET++ reproduces the measured $p_{\rm T}$-spectrum for central 
PbPb events at least up to eight orders of cross section magnitude. 
Multiplicities of the hard and soft components become comparable at 
$p_{\rm T} \sim 4.5$ GeV/$c$.

High transverse momentum region of hadron spectra ($p_{\rm T} > 5 \div 10$
GeV/$c$) is sensitive to parton production and jet quenching. 
In HYDJET++, they are determined by the PYQUEN parameters
for initial conditions of QGM formation~\cite{Lokhtin:2011qq}. The
nuclear modification factor $R_{\rm AA}$ is defined as a ratio of
particle yields in AA and pp collisions normalized on the mean
number of binary nucleon-nucleon sub-collisions $\left<N_{\rm coll}\right>$:
\begin{equation}
\label{raa}
R_{\rm AA}(p_{\rm T})=\frac{d^2N^{\rm AA}/d\eta dp_{\rm T}}
{\left<N_{\rm coll}\right>d^2N^{\rm pp}/d\eta dp_{\rm T}}~.
\end{equation}
In the absence of nuclear effects (in initial or final states) at
high $p_{\rm T}$, it should be $R_{\rm AA}=1$. In figure~\ref{fig2} (right), 
the calculated nuclear modification factor $R_{\rm AA}(p_{\rm T})$ for
charged hadrons ($|\eta|<1$) is compared with the CMS
data~\cite{CMS:2012aa} in 5\% of most central PbPb
collisions. The reasonable agreement with the data is achieved up to $p_{\rm T}
\sim 100$ GeV/$c$. However the simulated $p_{\rm T}$-dependence of $R_{\rm AA}$ 
looks rather weaker than the tendency of the data.  
It could indicate that the real energy dependence of partonic energy loss 
is somewhat weaker than it is expected from the model simulations. 
It is worth to remind here 
that the energy dependence of BDMS radiative loss~\cite{Baier:1999ds,Baier:2001qw} 
(the dominant mechanism of medium-induced energy loss in PYQUEN) is between 
$\Delta E \propto \ln {E}$ and $\Delta E \propto \sqrt {E}$ (depending on the 
energy of the radiated gluon and on the properties of the medium). 

\subsection{Femtoscopic momentum correlations}

Effects of quantum statistics and final state interactions make
the momentum (HBT) correlation functions of two or more particles
in their c.m.s. sensitive to space-time characteristics of
production process on the level of femtometers at small relative
momenta. Since HYDJET++ specifies the space-time structure of a
hadron emission source (for both soft and hard components), the
momentum correlation function can be introduced by the standard
weighting procedure~\cite{Podgoretsky:1982xu,Pratt:1984su}. Having
the information about final particle 4-momenta $p_i$ and
4-coordinates $x_i$ of emission points, we can calculate the
correlation function with the help of the weight procedure,
assigning a weight to a given particle combination and accounting
for effects of quantum statistics:
\begin{equation}
      w = 1 + \cos(q \cdot \Delta x),
\label{cf4}
\end{equation}
where  $q = p_1 -p_2 $ and $\Delta x = x_1 - x_2$. Then the
correlation function is defined as a ratio of a weighted histogram
of pair kinematic variables to an unweighted one. The
corresponding correlation widths are parametrized in terms of the
Gaussian correlation radii $R_i$ as follows:
\begin{eqnarray}
CF(p_{1},p_{2})= 1+\lambda\exp(-R_\mathrm{out}^2q_\mathrm{out}^2
-R_\mathrm{side}^2q_\mathrm{side}^2- \\ \nonumber
R_\mathrm{long}^2q_\mathrm{long}^2) \label{cf3}~,
\end{eqnarray}
where
$\mathbf{q}=(q_\mathrm{out},q_\mathrm{side},q_\mathrm{long})$ is
the relative three-mo\-men\-tum vector of two identical particles and
$\lambda$ is the correlation strength. The ``out'' axis is pointing 
along the pair transverse momentum, the ``side'' axis is perpendicular 
to the ``out'' direction in the transverse plane, and the ``long'' 
axis is along the beam.

\begin{figure*}
\vspace*{5cm}
\begin{center}
\resizebox{1.\textwidth}{!}{%
\includegraphics{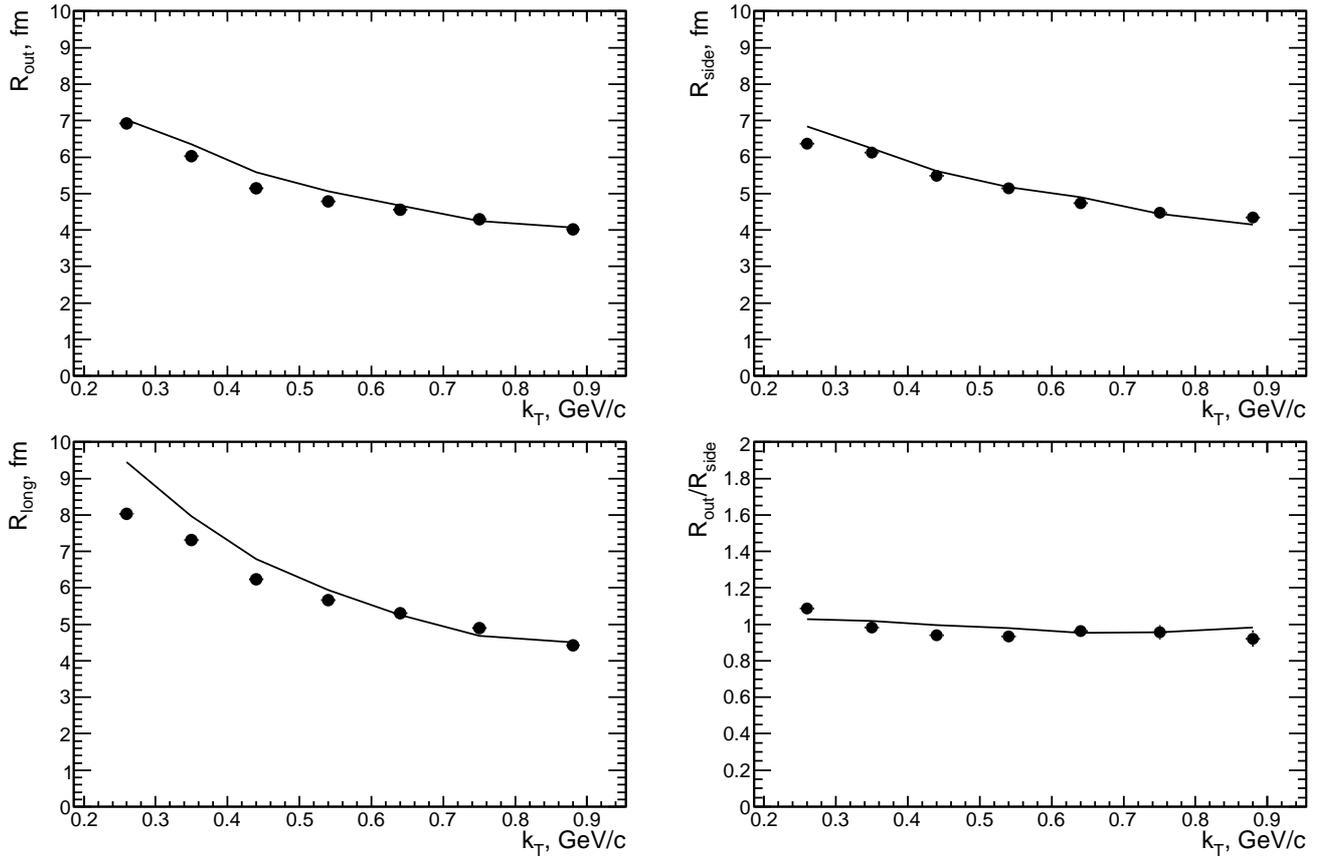}
}
\end{center}
\caption{$\pi^\pm \pi^\pm$ correlation radii as functions of pion pair 
transverse momentum $k_{\rm T}$ in 5\% of most central PbPb collisions at
$\sqrt s_{\rm NN}=2.76$ TeV. The points are the ALICE
data~\cite{Aamodt:2011mr}, curves are the simulated HYDJET++
events.} \label{fig3}
\end{figure*}

Figure~\ref{fig3} shows the fitted correlation radii as a function of mean 
transverse momentum $k_{\rm T}$ of pion pair ($|\eta|<0.8$) in 5\% of most 
central PbPb collisions at $\sqrt s_{\rm NN}=2.76$ TeV. The results of the 
HYDJET++ simulation are compared with the ALICE data~\cite{Aamodt:2011mr}.
Taking into account both soft and hard components allows us to reproduce
the measured $k_{\rm T}$-dependencies of the correlation radii.
 
Let us comment how our relatively simple model without detailed fluid dynamic evolution 
and hadronic cascade is able to fit the experimental data on the femtoscopic observables. 
In fact, there are 3D-hydrodynamics models which reproduce HBT radius 
parameters in heavy ion 
collisions at RHIC and LHC energies, e.g.~\cite{Sinyukov:2010yf,Bozek:2011ua}. 
A successful simultaneous description of hadronic yields and spectra, elliptic 
flow and correlation radii at RHIC and LHC is achieved also in hydro-kinetic 
model~\cite{Karpenko:2012yf}. These approaches include the hydrodynamical evolution 
and hadronic cascade calculations which are very computer time consuming. 
HYDJET++ is the multi-purpose event generator for a fast simulation (needed e.g. in Monte-Carlo studies 
of the experimental setup), and so it does not contain detailed hydrodynamical evolution and hadronic 
cascade. The treatment of the soft component in HYDJET++ is based on 
the fast simulation of the freeze-out hypersurface realized in FAST MC 
model~\cite{Amelin:2006qe,Amelin:2007ic} describing well 
the femtoscopic radii at RHIC energies. The main feature of this approach  
is that the hadron system expands hydrodynamically 
with the frozen chemical composition, cools down, and finally decays at some thermal
freeze-out hypersurface. The chemical potentials are fixed from the
experimental particle number ratios at the chemical freeze-out. 
In spite of lack of the hydrodynamical evolution in HYDJET++, the assumption of the 
conservation of the particle number ratios from the chemical to thermal freeze-out 
allows us to calculate the chemical potentials at the thermal freeze-out.
The absolute values of particle densities are determined by 
the free parameter of the model, effective pion chemical potential
$\mu_{\pi}^{\rm th}$ at the thermal freeze-out. The best tune of the temperatures 
at chemical and thermal freeze-out hypersurfaces and volume parameters allow us to 
reproduce simultaneously the data on the total multiplicity, hadronic spectra and correlation radii. At the 
same time, low momentum particles (with $p_{\rm T}<1$~GeV/$c$) coming from the hard component are 
emitted from the same thermal freeze-out hypersurface as soft component hadrons (see subsection 2.2).
 
\subsection{Elliptic flow}

\begin{figure*}
\vspace*{5cm}
\begin{center}
\resizebox{1.\textwidth}{!}{%
\includegraphics{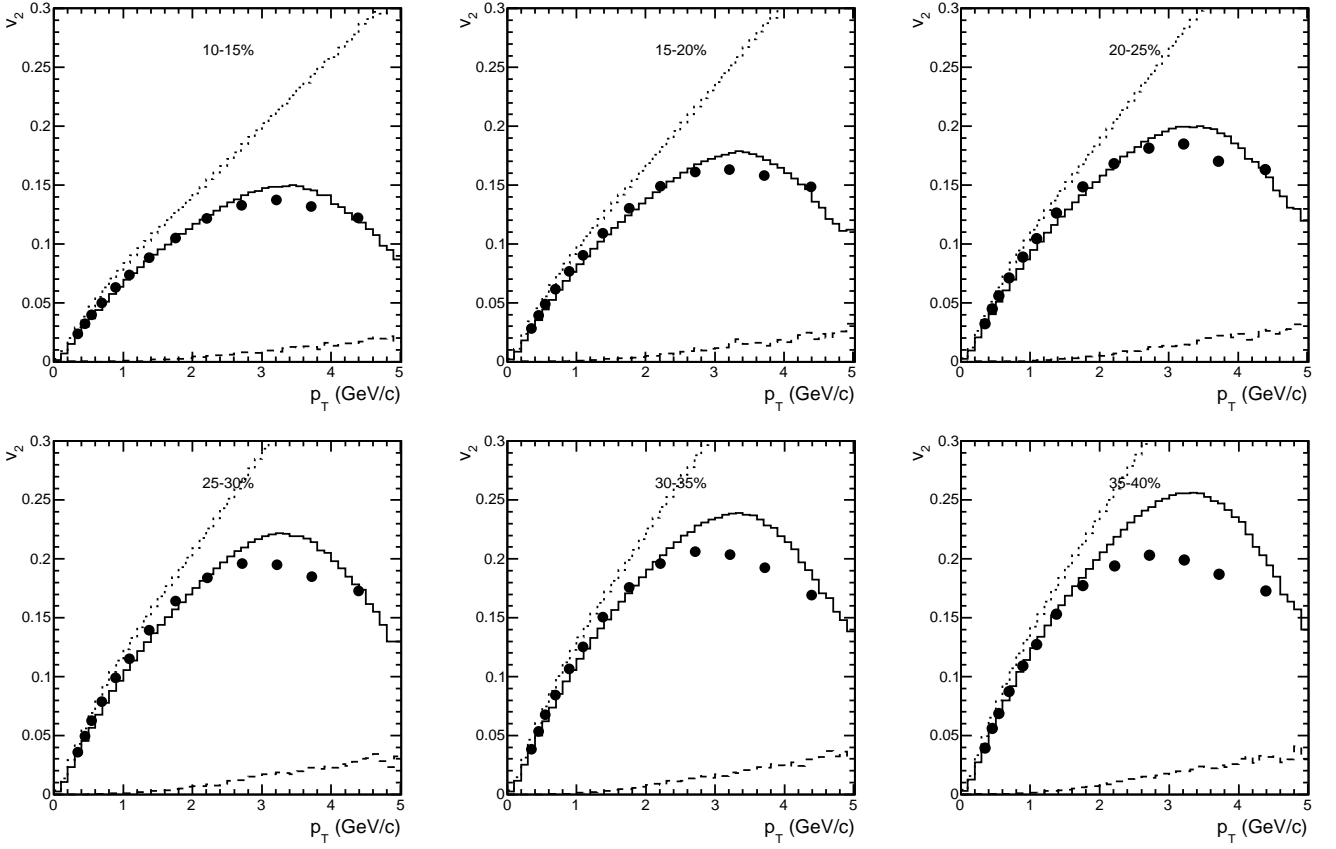}
}
\end{center}
\caption{Elliptic flow coefficient $v_2(p_{\rm T})$ for charged hadrons
at mid-rapidity ($|\eta|<0.8$) for different centralities of PbPb
collisions at $\sqrt s_{\rm NN}=2.76$ TeV. The points are the CMS
data~\cite{Chatrchyan:2012ta}, histograms are the simulated
HYDJET++ events (solid -- total result, dashed -- hard component,
dotted -- soft component).} 
\label{fig4}
\end{figure*}

\begin{figure*}
\vspace*{5cm}
\begin{center}
\resizebox{1.\textwidth}{!}{%
\includegraphics{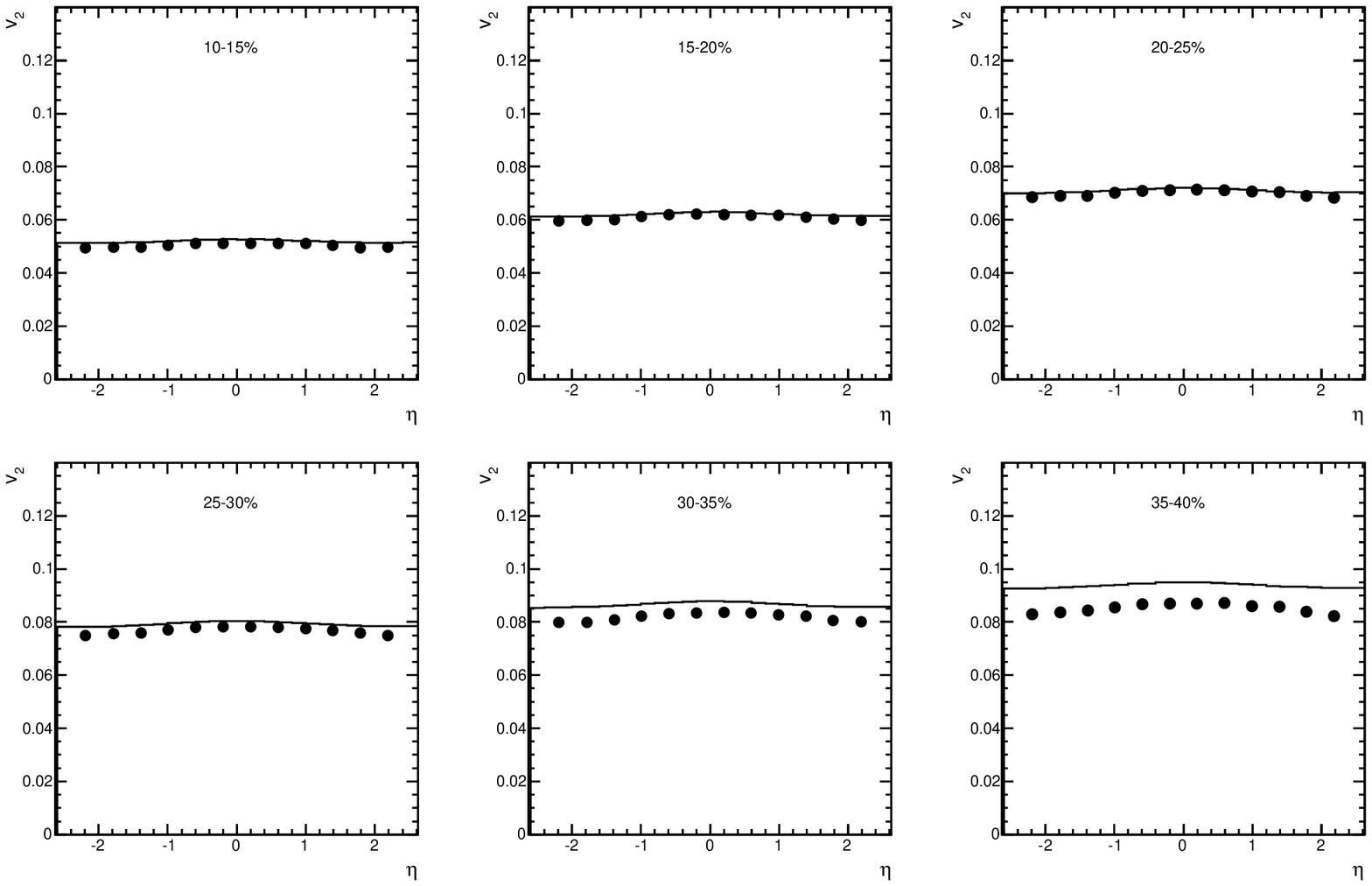}
}
\end{center}
\caption{Elliptic flow coefficient $v_2(\eta)$ ($0.3 < p_T < 3$ GeV/$c$) for charged hadrons at different centralities of PbPb collisions at $\sqrt s_{\rm NN}=2.76$ TeV. 
The points are the CMS data~\cite{Chatrchyan:2012ta}, histograms are the simulated
HYDJET++ events.} 
\label{fig5}
\end{figure*}

The elliptic flow coefficient $v_2$ is defined as the second-order
Fourier coefficient in the hadron distribution over the azimuthal
angle $\varphi$ relative to the reaction plane angle $\psi_{\rm R}$, so
that $v_2 \equiv \left< \cos{2(\varphi-\psi_{\rm R})} \right>$. It is an
important characteristic of physics dynamics at early stages of
non-central heavy ion collisions. According to the typical
hydrodynamic scenario, values of $v_2(p_{\rm T})$ at 
low-$p_{\rm T}$ are
determined mainly by internal pressure gradients of an expanding
fireball during the initial high density phase of the reaction,
while the elliptic flow at high-$p_{\rm T}$ is generated due to partonic
energy loss in an azimuthally asymmetric volume of QGM.
Figures~\ref{fig4} and \ref{fig5} show elliptic flow coefficient
$v_2$ as a function of the hadron transverse momentum $p_{\rm T}$ and
pseudorapidity $\eta$, respectively. The results of the HYDJET++
simulation are compared with the CMS data obtained from the 4-particle 
cumulant method~\cite{Chatrchyan:2012ta}. HYDJET++ reproduces 
$\eta$-dependence and $p_{\rm T}$ dependence of $v_2$ at mid-rapidity (up
to $p_{\rm T} \sim 5$ GeV/$c$ and $40$\% centrality). As it is
expected, taking into account the hard component changes the
hydrodynamical growth of $v_2$ with $p_{\rm T}$, so $v_2$ decreases at
$p_{\rm T} > 3\div 4$ GeV/$c$. The flat $\eta$-dependence of $v_2$ 
manifests the strong longitudinal flow effect taken into account in HYDJET++. 

\section{Conclusions}

The LHC data on multiplicity, hadron spectra, elliptic flow
and femtoscopic momentum correlations in PbPb collisions at $\sqrt
s_{\rm NN}=2.76$ TeV are analyzed in the framework of the HYDJET++
model. The significant influence of the jet production mechanism
on these observables is shown. We have estimated that the
contribution of the hard component to the total multiplicity is
about $25$\% at mid-rapidity in most central collisions, and
it decreases for more peripheral events.

Taking into account both hard and soft components and tuning input
parameters allow HYDJET++ to reproduce the LHC data on centrality and pseudorapidity 
dependence of charged particle multiplicity, $p_{\rm T}$-spectra, nuclear modification factor 
(up to $p_{\rm T} \sim 100$ GeV/$c$) and $\pi^\pm \pi^\pm$ correlation radii in central PbPb collisions, 
and $p_{\rm T}$- and $\eta$-dependencies of the elliptic flow coefficient $v_2$ (up to
$p_{\rm T} \sim 5$ GeV/$c$ and $40$\% centrality).

\begin{acknowledgement}
Discussions with I.~Arsene, L.~Bravina, D.~d'Enterria, A.~Gribu\-shin, V.~Korotkikh,  L.~Sarycheva, Yu.~Sinyukov, K.~Tiwonyuk and E.~Zabrodin are gratefully acknowledged.
We thank colleagues from CMS and ALICE collaborations for
fruitful cooperation. This work was supported by Russian Foundation for Basic Research (grants 10-02-93118 and 12-02-91505), Grant of President of Russian Federation for scientific Schools
supporting (3920.2012.2) and Dynasty Foundation.
\end{acknowledgement}

\begin{thebibliography}{99}

\bibitem{d'Enterria:2006su} D.~d'Enterria, J. Phys. {\bf G 34}, (2007) S53
\bibitem{BraunMunzinger:2007zz} P.~Braun-Munzinger and J.~Stachel, Nature
{\bf 448}, (2007) 302
\bibitem{Salgado:2009jp} C.~Salgado, Proceedings of European School of High-Energy Physics (2008) 239,  arXiv:0907.1219 [hep-ph]
\bibitem{Dremin:2010jx} I.M.~Dremin, A.V.~Leonidov, Phys. Usp. {\bf 53}, (2011) 1123
\bibitem{Aamodt:2010pb} K.~Aamodt, et al. (ALICE Collaboration), Phys. Rev. Lett. {\bf 105}, (2010) 252301
\bibitem{Aamodt:2010pa} K.~Aamodt, et al. (ALICE Collaboration), Phys. Rev. Lett. {\bf 105}, (2010) 252302
\bibitem{Aamodt:2010jd} K.~Aamodt, et al. (ALICE Collaboration), Phys. Lett. {\bf B 696}, (2011) 30
\bibitem{Aamodt:2010cz} K.~Aamodt, et al. (ALICE Collaboration), Phys. Rev. Lett. {\bf 106}, (2011) 032301
\bibitem{Aamodt:2011mr} K.~Aamodt, et al. (ALICE Collaboration), Phys. Lett. {\bf B 696}, (2011) 328
\bibitem{ALICE:2011ab} K.~Aamodt, et al. (ALICE Collaboration), Phys. Rev. Lett. {\bf 107}, (2011) 032301
\bibitem{Aamodt:2011by} K.~Aamodt, et al. (ALICE Collaboration), Phys. Lett. {\bf B 708}, (2012) 249 
\bibitem{Aamodt:2011vg} K.~Aamodt, et al. (ALICE Collaboration), Phys. Rev. Lett. {\bf 108}, (2012) 092301
\bibitem{Abelev:2012ej} B.~Abelev, et al. (ALICE Collaboration), JHEP {\bf 1203}, (2012) 053
\bibitem{Abelev:2012rv} B.~Abelev, et al. (ALICE Collaboration), arXiv:1202.1383 [hep-ex]
\bibitem{Abelev:2012nj} B.~Abelev, et al. (ALICE Collaboration), arXiv:1203.2160 [nucl-ex]
\bibitem{ALICE:2012aa} B.~Abelev, et al. (ALICE Collaboration), arXiv:1203.2436 [nucl-ex]
\bibitem{ALICE:2012di}  B.~Abelev, et al. (ALICE Collaboration), arXiv:1205.5761 [nucl-ex]
\bibitem{Aad:2010bu} G.~Aad, et al. (ATLAS Collaboration), Phys. Rev. Lett. {\bf 105}, (2010) 252303
\bibitem{Aad:2010px} G.~Aad, et al. (ATLAS Collaboration), Phys. Lett. {\bf B 697}, (2011) 294
\bibitem{ATLAS:2011yk} G.~Aad, et al. (ATLAS Collaboration), Phys. Lett. {\bf B 707}, (2012) 330
\bibitem{ATLAS:2011ag} G.~Aad, et al. (ATLAS Collaboration), Phys. Lett. {\bf B 710}, (2012) 363
\bibitem{Aad:2012bu} G.~Aad, et al. (ATLAS Collaboration), arXiv:1203.3087 [hep-ex]
\bibitem{Chatrchyan:2011sx} S.~Chatrchyan, et al. (CMS Collaboration), Phys. Rev. {\bf C 84}, (2011) 024906
\bibitem{Chatrchyan:2011ua} S.~Chatrchyan, et al. (CMS Collaboration), Phys. Rev.  Lett. {\bf 106}, (2011) 212301
\bibitem{Chatrchyan:2011ek} S.~Chatrchyan, et al. (CMS Collaboration), JHEP {\bf 1107}, (2011) 076
\bibitem{Chatrchyan:2011pe} S.~Chatrchyan, et al. (CMS Collaboration), Phys. Rev. Lett. {\bf 107}, (2011) 052302
\bibitem{Chatrchyan:2011pb} S.~Chatrchyan, et al. (CMS Collaboration), JHEP {\bf 1108}, (2011) 141
\bibitem{Chatrchyan:2012vq} S.~Chatrchyan, et al. (CMS Collaboration), Phys. Lett. {\bf B 710}, (2012) 256
\bibitem{CMS:2012aa} S.~Chatrchyan, et al. (CMS Collaboration),  Eur. Phys. J. {\bf C 72}, (2012) 1945
\bibitem{Chatrchyan:2012np} S.~Chatrchyan, et al. (CMS Collaboration), JHEP {\bf 1205}, (2012) 063
\bibitem{Chatrchyan:2012ni} S.~Chatrchyan, et al. (CMS Collaboration), Phys. Lett. {\bf B 712}, (2012) 176
\bibitem{Chatrchyan:2012wg} S.~Chatrchyan, et al. (CMS Collaboration), arXiv:1201.3158 [nucl-ex]
\bibitem{Chatrchyan:2012ta} S.~Chatrchyan, et al. (CMS Collaboration), arXiv:1204.1409 [nucl-ex]
\bibitem{Chatrchyan:2012xq} S.~Chatrchyan, et al. (CMS Collaboration), arXiv:1204.1850 [nucl-ex]
\bibitem{Chatrchyan:2012gt} S.~Chatrchyan, et al. (CMS Collaboration), arXiv:1205.0206 [nucl-ex]
\bibitem{Chatrchyan:2012mb} S.~Chatrchyan, et al. (CMS Collaboration), arXiv:1205.2488 [nucl-ex]
\bibitem{Muller:2012zq} B.~Muller, J.~Schukraft, B.~Wyslouch, arXiv:1202.3233 [hep-ex]
\bibitem{Lokhtin:2008xi} I.P.~Lokhtin, L.V~Malinina, S.V.~Petrushanko, A.M.~Snigirev,
I.~Arsene, K.~Tywoniuk, Comput. Phys. Commun. {\bf 180}, (2009) 779
\bibitem{Werner:2012xh} K.~Werner, Iu.~Karpenko, M.~Bleicher, T.~Pierog, S.~Porteboeuf-Houssais, arXiv:1203.5704 [nucl-th]
\bibitem{Lokhtin:2005px} I.P.~Lokhtin, A.M.~Snigirev, Eur. Phys. J. {\bf C 45}, (2006) 211
\bibitem{Amelin:2006qe} N.S.~Amelin et al., Phys. Rev. {\bf C 74}, (2006) 064901
\bibitem{Amelin:2007ic} N.S.~Amelin et al., Phys. Rev. {\bf C 77}, (2008) 014903
\bibitem{Wiedemann:1997cr} U.~Wiedemann, Phys. Rev. {\bf C 57}, (1998) 266
\bibitem{Torrieri:2004zz} G.~Torrieri, S.~Steinke, W.~Broniowski, W.~Florkowski,
J.~Letessier, J.~Rafelski, Comput. Phys. Commun. {\bf 167}, (2005) 229
\bibitem{Baier:1999ds}  R.~Baier, Yu. L.~Dokshitzer, A.H.~Mueller, D.~Schiff,
Phys. Rev. {\bf C 60}, (1999) 064902
\bibitem{Baier:2001qw} R.~Baier, Yu. L.~Dokshitzer, A.H.~Mueller, D.~Schiff,
Phys. Rev. {\bf C 64}, (2001) 057902
\bibitem{Dokshitzer:2001zm} Yu.L.~Dokshitzer, D.~Kharzeev, Phys. Lett. {\bf B 519}, (2001) 199
\bibitem{Bjorken:1982tu} J.D.~Bjorken, Fermilab publication Pub-82/29-THY, 1982
\bibitem{Braaten:1991jj} E.~Braaten, M.~Thoma, Phys. Rev. {\bf D 44}, (1991) 1298
\bibitem{Lokhtin:2000wm} I.P.~Lokhtin, A.M.~Snigirev, Eur. Phys. J.
{\bf C 16}, (2000) 527
\bibitem{Lokhtin:2011qq} I.P.~Lokhtin, A.V.~Belyaev, A.M.~Snigirev, Eur. Phys. J.
{\bf C 71}, (2011) 1650
\bibitem{Sjostrand:2006za} T.~Sjostrand, S.~Mrenna, P.~Skands, JHEP {\bf 0605}, (2006) 026
\bibitem{Chatrchyan:2011av} S.~Chatrchyan, et al. (CMS Collaboration),
JHEP {\bf 1108}, (2011) 086
\bibitem{Tywoniuk:2007xy} K.~Tywoniuk, I.C.~Arsene, L.~Bravina, A.B.~Kaidalov,
E.~Zabrodin, Phys. Lett. {\bf B 657}, (2007) 170
\bibitem{Paic:2005cx} G.~Paic, P.K.~Skowronski, J.Phys. {\bf G 31}, (2005) 1045
\bibitem{Floris:2011ru} M.~Floris (for the ALICE Collaboration), J. Phys. {\bf G 38},
(2011) 124025
\bibitem{Podgoretsky:1982xu} M.I.~Podgoretsky, Sov. J. Nucl. Phys. {\bf 37}, (1983) 272
\bibitem{Pratt:1984su} S.~Pratt, Phys. Rev. Lett. {\bf 53}, (1984) 1219
\bibitem{Sinyukov:2010yf} Yu.~Karpenko, Yu.~Sinyukov, Phys.Lett. {\bf B 688}, (2010) 50
\bibitem{Bozek:2011ua} P.~Bozek, Phys. Rev. {\bf C 85}, (2012) 034901
\bibitem{Karpenko:2012yf} I.A.~Karpenko, Y.M.~Sinyukov, K.~Werner, arXiv:1204.5351 [nucl-th]
\end{thebibliography}
\end{document}